\documentclass[11pt,letterpaper]{article}
\usepackage[utf8]{inputenc}
\usepackage{graphicx}
\usepackage{hyperref}
\usepackage{amsmath}
\usepackage{amssymb}
\usepackage[margin=1in]{geometry}
\usepackage{natbib}
\providecommand{\tightlist}{%
  \setlength{\itemsep}{0pt}\setlength{\parskip}{0pt}}

\title{Core Mondrian: Basic Mondrian beyond k-anonymity\thanks{Any mistakes in this paper are the sole responsibility of Airbnb. In the interest of encouraging others to adopt this methodology, Airbnb and the authors formally disavow any intent to enforce their copyright in the content of this technical paper. Airbnb will not be liable for any indemnification of claims of intellectual property infringement made against users by third parties. The authors acknowledge the use of ChatGPT, Claude, Gemini, and O3 models for assistance with code modification, code review, literature review, technical writing and editing. We would like to thank Dr. Latanya Sweeney (Data Privacy Lab at Harvard) for early, confidential review.}}
\author{
Adam Bloomston\thanks{Corresponding Author; they may be reached at antidiscrimination-papers@airbnb.com.}, Elizabeth Burke, Megan Cacace, Anne Diaz, \\
Wren Dougherty, Matthew Gonzalez, Remington Gregg, Yeliz Güngör, \\
Bryce Hayes, Eeway Hsu, Oron Israeli, Heesoo Kim, \\
Sara Kwasnick, Joanne Lacsina, Demma Rosa Rodriguez, Adam Schiller, \\
Whitney Schumacher, Jessica Simon, Maggie Tang, Skyler Wharton, \\
Marilyn Wilcken
}
\date{}

\begin{document}
\maketitle

\begin{abstract}
We present Core Mondrian, a scalable extension of the Original Mondrian
partition-based anonymization algorithm. A modular strategy layer
supports k-anonymity, allowing new privacy models to be added easily. A
hybrid recursive/queue execution engine exploits multi-core parallelism
while maintaining deterministic output. Utility-preserving enhancements
include NaN-pattern pre-partitioning, metric-driven cut scoring, and
dynamic suppression budget management. Experiments on the 48k-record UCI
ADULT dataset and synthetically scaled versions up to 1M records achieve
lower Discernibility Metric scores than Original Mondrian for numeric
quasi-identifier sets while parallel processing delivers up to 4×
speedup vs. sequential Core Mondrian. Core Mondrian enables
privacy-compliant equity analytics at production scale.
\end{abstract}

\section{\texorpdfstring{{Introduction}}{Introduction}}\label{h.kq4lu8f92ybs}

{Project Lighthouse, }{a tool Airbnb developed in partnership with
leading civil rights and privacy organizations, enables Airbnb to
uncover and address potential disparities in how users of different
perceived races may experience Airbnb, while at at the same time
}{maintaining privacy through k-anonymity and p-sensitive k-anonymity
\cite{Airbnb2024,Basuetal2020}. Each quasi-identifier (QID)
must appear in equivalence classes (groups of records that share
identical generalized quasi-identifier values) \cite{Samarati1998,Sweeney2002,Lietal2007} of size $\geq$ k, with each class containing at least
p distinct sensitive attribute values \cite{Truta2006} to prevent both
identity and sensitive attribute disclosure.}

{}

{Existing global-recoding tools (which apply single generalizations
across entire columns), though essential for
Lighthouse\textquotesingle s early development, failed to meet stringent
data quality requirements as datasets for analysis become more complex
over time. Core Mondrian addresses these limitations through
partition-based anonymization (which applies different generalizations
within each partition) that preserves analytical utility while meeting
privacy requirements.}

{}

{Core Mondrian provides three key contributions:}

{}

\begin{enumerate}
\tightlist
\item
  \textbf{Extensible Strategy Pattern architecture}~supporting multiple
  privacy models (see Section 3.4)
\item
  \textbf{Hybrid recursive-queue execution}~with multi-core parallelism (see
  Sections 3.5, 4.3)
\item
  \textbf{Data quality preserving features}~including NaN-pattern
  pre-partitioning and dynamic suppression budget management (see
  Sections 3.2, 3.6, 4.2)
\end{enumerate}

{}

{Before detailing our approach, we first examine existing solutions and
their limitations. Section 2 reviews related work, Section 3 details the
algorithm design, Section 4 presents experimental evaluation, and
Section 5 concludes with future work directions.}

\section{\texorpdfstring{{Related Work}}{Related Work}}\label{h.ow9zc171oaty}

{ARX \cite{Prasseretal2020} was invaluable during
Lighthouse\textquotesingle s initial prototyping, providing a mature
framework with comprehensive privacy models and utility metrics.
However, Airbnb\textquotesingle s specific production-scale requirements
and infrastructure constraints motivated the development of a custom
solution tailored to Project Lighthouse\textquotesingle s needs. The
utility degradation motivated the initial testing of Mondrian, and
eventually to the development of Core Mondrian, employing local
partition-based generalization to better preserve utility while
achieving the throughput and extensibility required for production
equity analytics.}

{}

{The Mondrian algorithm provides a local-recoding alternative to global
approaches through recursive, partition-based generalization. The
algorithm recursively divides datasets into smaller partitions,
selecting the quasi-identifier (QID) with the largest normalized range
and splitting at the median value. Each resulting partition must contain
at least k records to satisfy k-anonymity constraints.
Mondrian\textquotesingle s local partitioning approach better preserves
utility than global recoding, particularly for heterogeneous data
\cite{LeFevreetal2006}.}

{}

{However, several limitations motivated Core Mondrian\textquotesingle s
development: including lack of parallelization and minimal suppression
control. Researchers have proposed numerous extensions to address these
limitations: distributed processing approaches for scalability, and
utility-based anonymization with local recoding \cite{Xuetal2006} for
better partitioning strategies.}

{}

{While these extensions offer valuable improvements, they were
individually or collectively insufficient for Project
Lighthouse\textquotesingle s unique requirements: processing
Airbnb\textquotesingle s large-scale datasets with acceptable
performance, maintaining high data utility for nuanced equity
measurements using tailored metrics like Revised Information Loss Metric
(RILM), a revision and extension of the Information Loss Metric (ILM)
\cite{Byunetal2006,bloomstonetal2025}, robust handling of missing values, and
an extensible architecture supporting k-anonymity variants within
Python-based infrastructure. This gap necessitated Core Mondrian, which
integrates architectural and algorithmic enhancements specifically
engineered for Project Lighthouse.}

{}

{Having identified the limitations of existing approaches, we now
present Core Mondrian\textquotesingle s design and implementation to
address these challenges.}

\section{\texorpdfstring{{Core Mondrian Algorithm}}{Core Mondrian Algorithm}}\label{h.dyr84wa6b9wg}

\subsection{\texorpdfstring{{Algorithmic Overview}}{Algorithmic Overview}}\label{h.cg165djf5x7o}

{Core Mondrian builds on Mondrian\textquotesingle s local-partitioning
approach but introduces five interlocking mechanisms that together
constitute the Core Mondrian engine. These components operate
concurrently and continuously influence one another throughout
anonymization:}

{}

{• }{Section 3.2 (Missing Value Handling)}{~-- NaN-pattern
pre-partitioning establishes how records sharing identical missing value
patterns are pre-grouped, creating fundamental partition boundaries that
persist throughout processing.}

{}

{• }{Section 3.3 (Partitioning Decision Process)}{~-- A multi-stage
funnel with utility-aware cut scoring systematically evaluates candidate
cuts using RILM-driven selection whenever any partition is considered
for splitting.}

{}

{• }{Section 3.4 (Extensible Architecture)}{~-- The Strategy Pattern
provides a pluggable framework where }{\texttt{Implementation\_Base}}{~decouples
privacy logic from the partitioning engine, supplying privacy models,
scoring functions, and cut policies to the decision process.}

{}

{• }{Section 3.5 (Hybrid Execution Model)}{~-- A runtime that
dynamically routes child partitions between immediate recursive
processing (for small partitions) and parallel queue-based processing
(for large partitions), enabling multi-core distribution of tasks.}

{}

{• }{Section 3.6 (Dynamic Suppression Budget Management)}{~-- A global
budget tracking system that monitors and constrains suppression
throughout all partitioning decisions, providing utility-aware control
of record removal.}

{}

{Together, these mechanisms form an integrated control loop: every
partition respects missing-value groupings (3.2), undergoes systematic
cut evaluation (3.3) using pluggable strategies (3.4), gets routed by
the hybrid scheduler (3.5), all while budget constraints (3.6) influence
every decision. The algorithm terminates when no further
budget-respecting splits are possible.}

\subsection{\texorpdfstring{{Missing Value Handling}}{Missing Value Handling}}\label{h.l6a0dbtxfrcr}

{A significant challenge in applying k-anonymity algorithms to
real-world datasets, such as those encountered in Project Lighthouse, is
the pervasive presence of \textbf{missing values}. Anonymization
implementations often lack explicit strategies for robustly handling
missing values.}

{}

\subsubsection{NaN-Pattern Pre-Partitioning}

{Core Mondrian, specifically designed to address the
practical data quality challenges inherent in datasets like those used
in Project Lighthouse, incorporates a NaN-pattern
pre-partitioning~mechanism. Prior to initiating the recursive
multidimensional partitioning process, the input dataset undergoes a
pre-emptive partitioning. This step is facilitated by a
}{\texttt{nan\_generator}}{~utility, which groups records into distinct initial
partitions. Each such partition comprises records that share an
identical pattern of missing (NaN) and present values across all
designated quasi-identifier (QID) attributes. For example, records where
}{\texttt{QID\_A}}{~is NaN and }{\texttt{QID\_B}}{~is present would form one initial
partition, separate from records where }{\texttt{QID\_A}}{~is present and
}{\texttt{QID\_B}}{~is NaN, or records where all QIDs are present, or where all
QIDs are NaN.}

{}

{These initial, NaN-pattern-homogeneous partitions are then
\textbf{independently processed}~by the Core Mondrian anonymization
algorithm. Technically, each of these initial partitions is represented
as a }{\texttt{Node\_DeferredCutData\_\_PartitionRoot}}{~object within the
}{\texttt{MondrianTree}}{~structure (see Section 3.5 for hybrid execution
details) and is processed as a separate root for a sub-anonymization
task. This pre-emptive partitioning guarantees that subsequent
generalization hierarchies and cut-selection heuristics operate on data
subsets with consistent missingness characteristics.}

\subsection{\texorpdfstring{{Partitioning Decision Process}}{Partitioning Decision Process}}\label{h.m15pz8uttz13}

{Core Mondrian systematically narrows the set of possible cuts at each
decision point through a multi-stage filtering process: rather than
selecting a partition split in a single step, the algorithm processes
candidates through a series of increasingly stringent criteria. This
process may be conceptualized as a Cut Funnel and provides insight into
algorithmic selectivity, computational bottlenecks, and potential areas
for optimization.}

{}

\subsubsection{Cut Funnel}

{The Cut Funnel~begins with all possible quasi-identifier (QID)
attributes that could potentially be split and progressively filters
candidates through multiple stages. While the term "funnel" suggests a
linear narrowing process, the actual implementation involves evaluation
at various stages, with multiple cuts potentially considered and
ultimately 0 or 1 cut selected as the winner. Each stage applies
different filtering criteria, from basic eligibility checks to
sophisticated utility scoring. Section 4.4 provides a concrete example
of this funnel in action on real data.}

{}

\subsubsection{Cut Choices}

{The initial generation of Cut Choices~differentiates Core Mondrian
from the original algorithm significantly. The original Mondrian splits
each partition along the quasi-identifier (QID) with the \textbf{largest
normalized range} \cite{LeFevreetal2006}. In contrast, Core Mondrian
delegates dimension selection to
}{\texttt{Implementation\_Base.cut\_choices()}}{~(see Section 3.4), allowing
different strategies appropriate for each privacy model implementation.
For k-anonymity, Core Mondrian uses an \textbf{RILM-based approach}~that is
similar in principle to the normalized range concept from the original
Mondrian paper with the following extensions:}

{}

{• Calculates current range (min, max) for numerical QIDs and selects
the smallest pre-calculated percentile-based domain that encompasses
this range, computing RILM scores to handle fat-tailed distributions
effectively\\
• Determines current generalized values using generalization hierarchies
and computes RILM scores accordingly for categorical QIDs\\
• Sorts all QIDs by RILM score (lower is better) to prioritize
dimensions with least expected information loss}

{}

\subsubsection{Dynamic Breakout}

{Dynamic breakout~occurs when a QID\textquotesingle s RILM gets too
high locally, causing it to be excluded from cut choice consideration so
that further partitioning focuses on attributes with more opportunity
for data quality improvement---an implementation-appropriate tradeoff
between privacy and data utility.}

{}

\subsubsection{Proposed Cut}

{Proposed Cut~generation represents another area of enhancement over
the original approach. The original Mondrian always splits numerical
QIDs at the median value for balanced partitioning \cite{LeFevreetal2006}
and relies on taxonomies for categorical QIDs, partitioning based on
generalization hierarchy. Core Mondrian supports both median cuts and
\textbf{histogram bin edges}~for numerical attributes:}

{}

\begin{itemize}
\item \textbf{Median cut mode:} Reproduces classic behavior
\item \textbf{Bin-edge mode:} Considers multiple split candidates based on data distribution (histogram bins) within the current partition
\end{itemize}

{}

{For categorical attributes, Core Mondrian uses generalization
hierarchies (generalization trees, or \textbf{gtrees}) where splits occur at
the immediate children of the lowest common ancestor (LCA) for the set
of observed values in the current partition.}

{}

\subsubsection{Final Cut}

{When multiple valid cuts are available, Core Mondrian employs
utility-aware scoring to further narrow down to a Final Cut. This
scoring mechanism evaluates proposed cuts based on how well they
preserve data distribution within partitions, using standard deviation
as the key metric. For both numerical and categorical QIDs, cuts that
reduce within-partition variance (lower standard deviation) are
preferred, as they indicate better data homogeneity and thus higher
utility preservation. The scoring returns values in the range {[}-1.0,
1.0{]}, where negative scores indicate improved data distribution and
are favored during selection. Finally a cut may be \textbf{Accepted}~if it
does not violate the suppression budget (see Section 3.6).}

\subsection{\texorpdfstring{{Extensible Architecture}}{Extensible Architecture}}\label{h.du3y9i3p5hgn}

{A cornerstone of Core Mondrian\textquotesingle s design, driven by the
complex and evolving privacy requirements of initiatives like Project
Lighthouse \cite{Basuetal2020}, is its extensible architecture. This
architecture was engineered to systematically decouple the core
multidimensional partitioning engine from the specific rules and
validation logic of any particular privacy model that relies on grouping
records into equivalence classes and employing generalization and
suppression techniques \cite{Samarati1998}. This separation of concerns
is critical for a production-grade system intended for long-term use,
allowing Core Mondrian to adapt to new privacy definitions, utility
metrics, and analytical needs without requiring modifications to its
fundamental data splitting, parallelism, or tree management mechanisms.}

{}

{The primary architectural pattern employed to achieve this flexibility
is the \textbf{Strategy Pattern} \cite{Gammaetal1995}. In this design, the
}{\texttt{CoreMondrian}}{~class acts as the \textbf{Context}, coordinating the overall
anonymization process, including managing the dataset, processing queue,
and tree construction. It delegates all privacy-specific
decision-making---such as determining partition validity, selecting
optimal cut points, and applying final generalization or
suppression---to an interchangeable \textbf{Strategy}~object. This strategy
object is an instance of a concrete class that implements a common
abstract base class, }{\texttt{Implementation\_Base}}{. An instance of a specific
privacy strategy is provided to }{\texttt{CoreMondrian}}{~at runtime, enabling a
"plug-and-play" approach for different privacy models and utility
optimization techniques.}

{}

{The }{\texttt{Implementation\_Base}}{~class (defined in }{\texttt{implementation.py}}{)
serves as the contract, or interface, for all privacy model strategies.
It defines a set of key methods that concrete implementations can
customize or override. These methods encapsulate the essential decision
points and logic required by the }{\texttt{CoreMondrian}}{~engine during the
recursive partitioning process, particularly the cut selection logic
detailed in Section 3.3.}

{}

{To extend Core Mondrian with a new privacy model, such as l-diversity
\cite{Machanavajjhalaetal2007} or t-closeness \cite{Lietal2007},
or to introduce a novel utility metric, a developer would create a new
Python class inheriting from }{\texttt{Implementation\_Base}}{. This new class
would override the necessary methods to reflect the validation rules,
cut preferences, generalization procedures, and suppression logic of the
new model or metric. The }{\texttt{CoreMondrian}}{~engine itself would require no
changes, readily accepting an instance of this new strategy.}

\subsection{\texorpdfstring{{Hybrid Execution Model}}{Hybrid Execution Model}}\label{h.5faednbblnqk}

{To address the scalability challenges inherent in processing
large-scale datasets, such as those encountered in Project Lighthouse
\cite{Basuetal2020}, Core Mondrian implements a \textbf{hybrid execution
model}. The Original Mondrian algorithm is recursively defined, but
that recursive stack isn\textquotesingle t required for solving
sub-problems. However, Core Mondrian does utilize the recursive stack
for Dynamic Suppression Budget Management (see Section 3.6).
Additionally, Core Mondrian utilizes a queue-based approach to
parallelize work across multiple cores. Therefore, Core Mondrian
implements a hybrid execution model~to strategically combine
recursive processing with a queue-based system to optimize resource
utilization and manage computational load.}

{}

{The partitioning process is primarily orchestrated by the
}{\texttt{CoreMondrian.anonymize}}{~method, which manages a central processing
queue, and the }{\texttt{CoreMondrian.make\_cut}}{~method, which contains the
core logic for splitting a partition. The execution strategy for a child
partition generated by }{\texttt{make\_cut}}{~is determined by its size relative
to a configurable threshold, }{\texttt{recursive\_partition\_size\_cutoff}}{.
This parameter dictates whether a partition is processed immediately via
a recursive call or deferred for later processing via the central
queue.}

{}

\subsubsection{Recursive Processing for Smaller Partitions}

{When a child partition\textquotesingle s record count is below the
}{\texttt{recursive\_partition\_size\_cutoff}}{, it is processed immediately
through recursive calls to }{\texttt{make\_cut}}{. This approach is efficient
for smaller partitions where the overhead of task scheduling and
parallel coordination would outweigh the benefits of parallelization.}

{}

\subsubsection{Queue-Based System for Larger Partitions}

{Conversely, child
partitions whose record count meets or exceeds the
}{\texttt{recursive\_partition\_size\_cutoff}}{~are deemed substantial enough
to benefit from deferred and potentially parallel processing. These
larger partitions are encapsulated as
}{\texttt{Node\_DeferredCutData}}{~objects and returned to the
}{\texttt{anonymize}}{~method. This method then adds these deferred tasks to a
central processing queue (implemented using }{\texttt{collections.deque}}{),
which serves as a pool of pending work. If parallel processing is
enabled, tasks from this queue are submitted as futures to a
}{\texttt{concurrent.futures.ProcessPoolExecutor}}{. This queue-based mechanism
is essential for scalability, as it:}

{}

\begin{itemize}
\tightlist
\item
  {Prevents excessively deep recursion stacks, mitigating risks of
  }{\texttt{StackOverflowError}}{~and high memory consumption for datasets that
  partition into many levels.}
\item
  {Enables efficient load balancing and parallel execution of
  computationally intensive partitioning tasks across multiple cores.}
\end{itemize}

{}

{When parallel processing is enabled, the main
process coordinates result collection as worker processes complete their
tasks. Each worker returns a }{\texttt{MondrianTree}}{~sub-tree representing
anonymization decisions for its assigned partition. The main process
retrieves completed results using
}{\texttt{concurrent.futures.as\_completed}}{~and stitches sub-trees into the
global tree structure using }{\texttt{MondrianTree.stitch\_in\_subtree()}}{. To
maintain deterministic output despite potentially out-of-order task
completion, suppression budgets are pre-calculated and allocated before
submitting tasks to the parallel pool.}

{}

{Partitions processed via the queue typically receive a fixed,
proportionally allocated suppression budget upfront, and dynamic
backtracking for budget overruns within that individual deferred task is
generally not supported at the global queue level. Initial partitions,
such as those derived from distinct NaN-patterns in quasi-identifiers
(QIDs) (see Section 3.2), are also managed through this central queue,
ensuring a unified approach to processing substantial data segments.}

{}

\subsubsection{Threshold Management and Optimization}

{The }{\texttt{recursive\_partition\_size\_cutoff}}{~parameter is critical for
optimizing the hybrid model\textquotesingle s performance. Its value
represents a trade-off: a low cutoff might lead to excessive overhead
from managing and parallelizing very small tasks, diminishing the
benefits of recursion. Conversely, a high cutoff might underutilize
parallel processing capabilities or lead to deeper-than-optimal
recursion for tasks that could have been efficiently parallelized.
Empirical tuning of this threshold, as discussed in Figure~\ref{fig:runtime-vs-cutoff}, is often
necessary to find an optimal balance for specific dataset
characteristics and hardware environments, thereby maximizing runtime
performance.}

{}

{In summary, Core Mondrian\textquotesingle s hybrid execution model
provides a robust and scalable partitioning engine. By dynamically
choosing between direct recursion for smaller tasks and a queue-based
system for larger ones, it optimizes performance, manages memory
effectively, and smoothly integrates with suppression budget management
(Section 3.6), making it well-suited for the demanding requirements of
anonymizing extensive datasets.}

\subsection{\texorpdfstring{{Dynamic Suppression Budget Management}}{Dynamic Suppression Budget Management}}\label{h.wn5mck8orln3}

{Balancing rigorous privacy guarantees with high data utility is a
central challenge for production anonymization systems, particularly in
high-stakes analytical contexts like Project Lighthouse. Excessive or
poorly managed \textbf{suppression}---the removal of records that cannot be
grouped without excessive generalization---not only degrades utility but
risks introducing bias, especially for underrepresented groups critical
to equity analyses.}

{}

{Core Mondrian introduces \textbf{dynamic suppression budget management}: a
principled, global-to-local framework for actively governing suppression
across the anonymization tree. This approach provides explicit,
auditable limits on suppression while preserving analytical signal.}

{}

\subsubsection{Global Budget Definition}

{}

{The global suppression budget S\_max is defined as S\_max = N · (1 -
p\_min) · multiplier, where N is the total number of records, p\_min $\in$
{[}0, 1{]} is the minimum proportion of records to retain (typically
0.98-0.99 for Project Lighthouse), and multiplier is an optional safety
margin factor. This cap maintains predictable, controlled record
retention across the anonymization process.}

{}

\subsubsection{Proportional Budget Allocation}

{}

{Core Mondrian computes S\_max and proportionally allocates this budget
during each partitioning step. When the dataset is split (e.g., by
missing value patterns, see Section 3.2), each initial partition P\_i
receives S\_i = \textbar P\_i\textbar{} · (S\_max/N). As partitions are
further split, any suppression incurred is first deducted, then
remaining budget is allocated using a two-phase approach:}

{}

\begin{itemize}
\item \textbf{Deferred partitions:} Queued partitions receive proportional budget allocation upfront
\item \textbf{Recursive partitions:} Processed in largest-to-smallest order, each receives full remaining budget
\end{itemize}

{}

{After processing each recursive child partition, the algorithm verifies
budget constraints and backtracks if S\_remaining \textless{} 0,
abandoning costly cut paths to seek alternatives with less suppression.}

{}

\subsubsection{Integration with Hybrid Processing}

{}

{Dynamic suppression budget management integrates tightly with Core
Mondrian\textquotesingle s hybrid execution model (Section 3.5). For
small partitions processed recursively, suppression budget is
\textbf{dynamically tracked}~and updated as each child node is processed. If
cumulative suppression exceeds allocation, the algorithm backtracks,
abandoning the current cut for a less costly alternative.}

{}

{Conversely, large partitions queued for parallel execution receive
\textbf{fixed budget allocation}~upfront. Once submitted to the processing
queue, their outcomes do not trigger parent-level backtracking, ensuring
scalability and deterministic reproducibility at the cost of reduced
cross-partition coordination.}

{}

{With Core Mondrian\textquotesingle s architecture and enhancements
detailed, we now evaluate its performance empirically against existing
approaches.}

\section{\texorpdfstring{{Experimental Evaluation}}{Experimental Evaluation}}\label{h.1of45og5mt6s}

\subsection{\texorpdfstring{{Experimental Setup}}{Experimental Setup}}\label{h.8wejotf87h3v}

{Dataset:}{~All experiments were conducted on the }{UCI ADULT Census
Income dataset \cite{Beckeretal1996}. The base dataset was loaded
using the }{\texttt{ucimlrepo}}{~library and preprocessed to join features and
target variables, standardize income formats, select only relevant
columns for analysis, replace missing values (represented as
\textquotesingle?\textquotesingle) with NaN, and remove non-QID columns
(}{\texttt{fnlwgt}}{, }{\texttt{education-num}}{). The columns included in the dataset
combine all QIDs used in the experiments plus demographic columns,
including numerical fields (age, capital-gain, capital-loss,
hours-per-week) and categorical fields (workclass, education,
marital-status, occupation, relationship, native-country, race, sex,
income).}

{}

{Synthetic Dataset Generation:}{~For scalability tests, we created
synthetically larger datasets using a Generative Adversarial Network
(GAN) approach, specifically the CTGAN implementation from the Synthetic
Data Vault (SDV) library \cite{Xuetal2019}. The CTGAN model was
trained on the base dataset for 1000 epochs, with the model
automatically detecting data types and distributions from the input
data. Once trained, the model generated synthetic datasets with scaling
factors of 2, 5, 10, 15, and 20 times the original size. For the size
factor 1, the original dataset was used as-is.}

{}

{Quasi-Identifier Sets:}{~We used several QID sets of varying
dimensionality:}

{}

\begin{itemize}
\tightlist
\item
  {1-4D Numeric QID Sets:}{~For information loss (Figures~\ref{fig:dm-rilm-comparison}, \ref{fig:privacy-utility-tradeoff}-\ref{fig:qid-dimensionality-impact}) and
  runtime (Figure~\ref{fig:runtime-comparison}) comparisons, we used numeric QID combinations of
  dimensions 1-4 (e.g., 1-QID: }{age}{~alone; 4-QID: all combinations of
  }{age}{, }{hours-per-week}{, }{capital-gain}{, and }{capital-loss}{).}
\item
  {Standard QID Sets:}{~For experiments focused on algorithm
  characteristics, we defined three standardized sets:}
\end{itemize}

\begin{itemize}
\tightlist
\item
  {4-QID:}{~}{age}{, }{race}{, }{sex}{, }{workclass}
\item
  {6-QID:}{~}{age}{, }{marital-status}{, }{occupation}{,
  }{native-country}{, }{sex}{, }{workclass}
\item
  {8-QID:}{~}{age}{, }{marital-status}{, }{occupation}{,
  }{native-country}{, }{sex}{, }{workclass}{, }{hours-per-week}{,
  }{capital-gain}
\end{itemize}

{}

{The dimensionality comparison (Figure~\ref{fig:qid-dimensionality-impact}) uses all three standard sets,
while most other experiments (Figures~\ref{fig:suppression-vs-cutoff}, \ref{fig:parallel-scalability}-\ref{fig:scalability-analysis}) use the 6-QID set as a
representative balance between dimensionality and computational
complexity.}

{}

{Algorithms:}

{}

\begin{itemize}
\tightlist
\item
  {Core Mondrian:}{~Our proposed implementation, featuring the
  extensible strategy pattern, hybrid execution model, parallel
  processing, NaN-pattern pre-partitioning, dynamic suppression budget
  management, and utility-aware cut scoring.}
\item
  {Original Mondrian (Baseline):}{~A faithful baseline implementation of
  the original Mondrian algorithm \cite{LeFevreetal2006}, constrained
  to single-threaded execution, median-cut partitioning, and no explicit
  mechanisms for suppression budget management or advanced NaN handling.
  Our baseline implementation focused on numeric-only quasi-identifiers
  (QIDs) for direct comparison, though the original algorithm supports
  categorical attributes via generalization hierarchies.}
\end{itemize}

{}

{While ARX was instrumental in early prototyping (Section 2), it was not
included in this direct comparison due to architectural and language
incompatibilities that made a side-by-side performance analysis on our
specific framework infeasible. Original Mondrian serves to isolate and
measure the impact of Core Mondrian\textquotesingle s specific
enhancements.}

{}

{Evaluation Metrics:}

{}

\begin{itemize}
\tightlist
\item
  {Discernibility Metric (DM):}{~An information loss metric that
  penalizes large equivalence classes. It is calculated as the sum of
  the squared sizes of all equivalence classes, with a higher penalty
  applied to suppressed records. A lower DM value indicates less
  information loss and is preferable \cite{Bayardoetal2005}.}
\item
  {Revised Information Loss Metric (RILM):}{~A utility metric that
  measures the preservation of data granularity on a scale from 0 to 1,
  where higher is better. It is calculated based on the range of
  generalized values relative to the total domain range for each QID
  \cite{bloomstonetal2025}.}
\item
  {Runtime:}{~Wall-clock time in seconds.}
\item
  {Suppression Rate:}{~The percentage of records suppressed from the
  original dataset.}
\end{itemize}

\subsection{\texorpdfstring{{Information Loss Analysis}}{Information Loss Analysis}}\label{h.30p6k1hnm6ql}

{We first evaluate Core Mondrian\textquotesingle s ability to preserve
data utility by minimizing information loss compared to the baseline.}

{}

\subsubsection{Core Mondrian vs. Original Mondrian}

{To provide a comprehensive comparison, we tested both algorithms across
numeric QID sets of increasing dimensionality (1-QID through 4-QID sets,
using combinations of }{age}{, }{hours-per-week}{, }{capital-gain}{, and
}{capital-loss}{) and a range of k values.}

{}

\begin{figure}[h]
\centering
\includegraphics[width=0.8\textwidth,keepaspectratio]{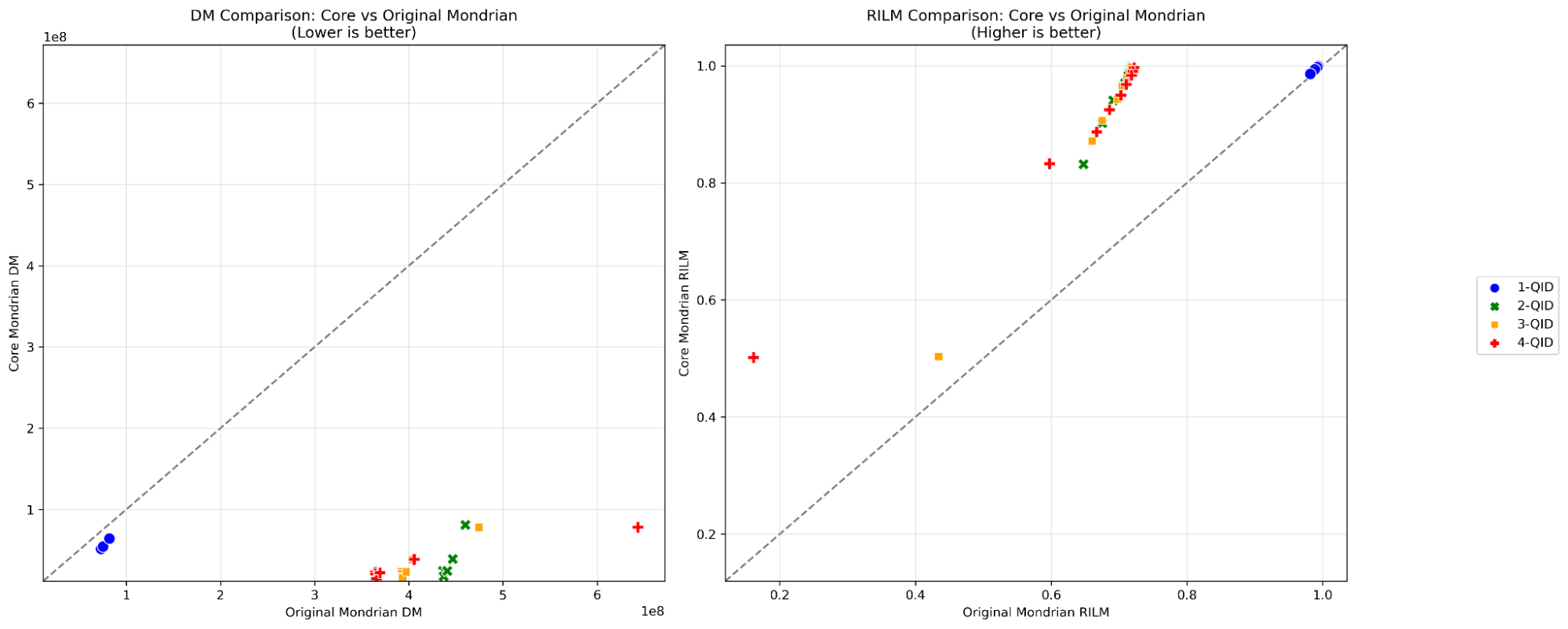}
\caption{Core Mondrian consistently achieves lower DM scores and higher RILM scores than Original Mondrian}
\label{fig:dm-rilm-comparison}
\end{figure}

{Figure~\ref{fig:dm-rilm-comparison} provides an aggregate comparison across all tested
configurations. The left panel plots the Discernibility Metric (DM) of
Core Mondrian against Original Mondrian. All points fall below the 1:1
reference line, indicating that Core Mondrian consistently achieves a
lower (better) DM score. The right panel plots the RILM scores, where
all points lie above the reference line, showing Core Mondrian preserves
higher utility. This advantage is maintained across all tested QID
dimensionalities \textgreater{} 2 and k values---a single QID shows
little discernible performance gap between Core and Original Mondrian.}

{}

\begin{figure}[h]
\centering
\includegraphics[width=0.8\textwidth,keepaspectratio]{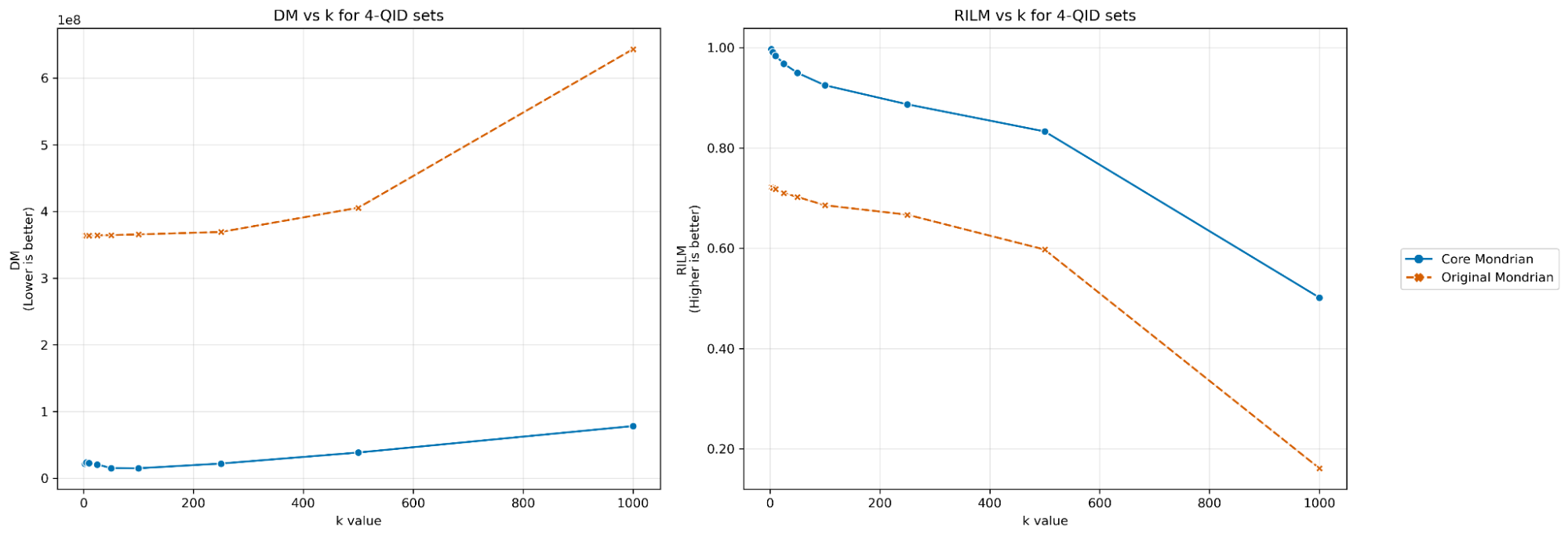}
\caption{Information loss vs k for 4-QID sets}
\label{fig:information-loss-vs-k}
\end{figure}

{Dimensional scaling effects become more apparent with higher numbers of
QIDs, as demonstrated in the 4-QID analysis (Figure~\ref{fig:information-loss-vs-k}). With a 4-QID
set, Core Mondrian\textquotesingle s DM increases moderately with k,
while Original Mondrian\textquotesingle s DM grows more steeply,
widening the performance gap. Concurrently, Core
Mondrian\textquotesingle s RILM remains significantly higher than
Original Mondrian\textquotesingle s across all values of k.}

{}

\subsubsection{Impact of k on Privacy-Utility Tradeoff}

{The choice of k directly controls the privacy-utility tradeoff. We
analyzed this relationship for Core Mondrian using the 4-QID set, which
is most relevant to practical applications and aligns with our runtime
analysis in Section 4.3.}

{}

\begin{figure}[h]
\centering
\includegraphics[width=0.8\textwidth,keepaspectratio]{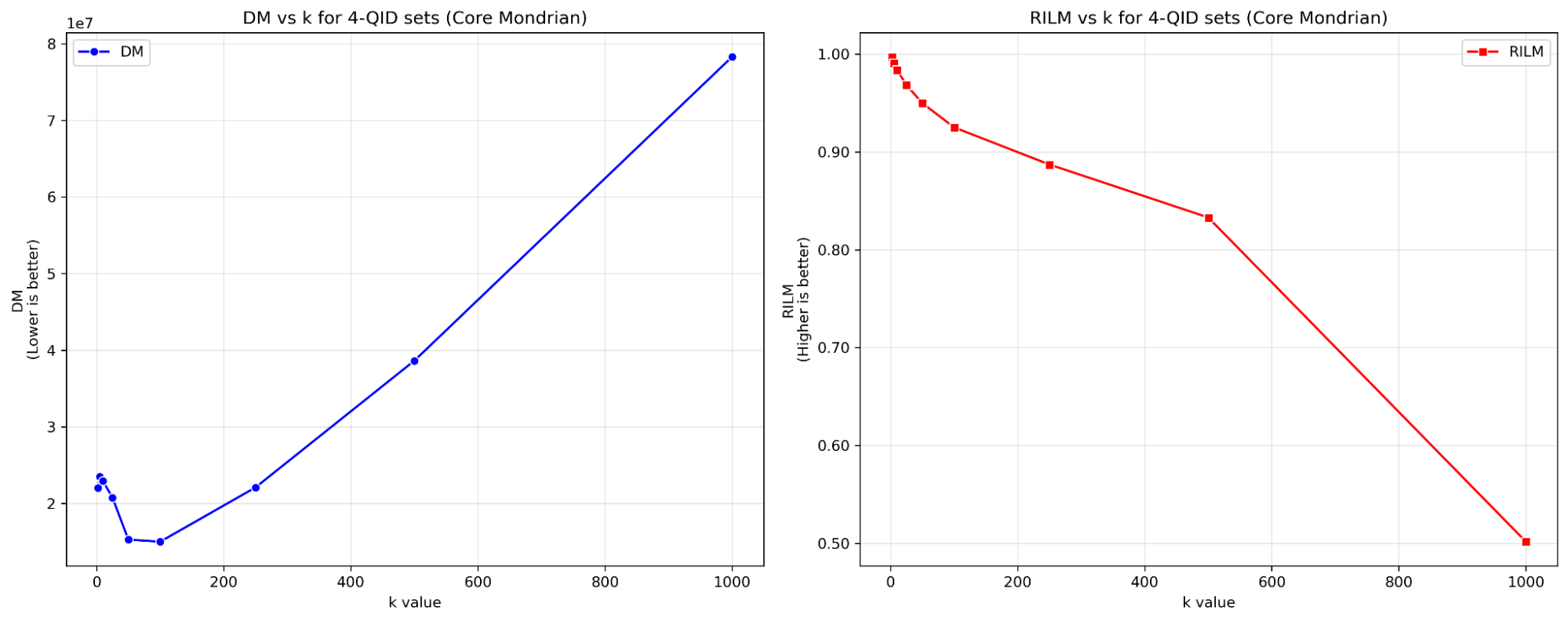}
\caption{Privacy-utility tradeoff showing DM and RILM vs k for 4-QID sets}
\label{fig:privacy-utility-tradeoff}
\end{figure}

{As shown in Figure~\ref{fig:privacy-utility-tradeoff}, increasing the privacy parameter k leads to a
clear tradeoff. The Discernibility Metric (DM) increases with k, rising
more steeply at higher values as equivalence classes must be made larger
to satisfy stricter privacy constraints. Correspondingly, the RILM score
gradually decreases from nearly perfect 1.00 at k=5, crossing 0.90 at
approximately k=200, and going as low as 0.50 for k=1000---signifying a
reduction in data granularity. This 4-QID analysis highlights the
fundamental cost of achieving stronger k-anonymity guarantees in
higher-dimensional data, which is particularly relevant for real-world
applications like Project Lighthouse where multiple attributes need
protection.}

{}

\subsubsection{Impact of QID Dimensionality}

{The "curse of dimensionality" is a known challenge in data
anonymization \cite{Aggarwal2005}. We evaluated Core
Mondrian\textquotesingle s performance using our three standard QID sets
(4-QID, 6-QID, and 8-QID) on the ADULT dataset for a fixed k=5.}

{}

\begin{figure}[h]
\centering
\includegraphics[width=0.8\textwidth,keepaspectratio]{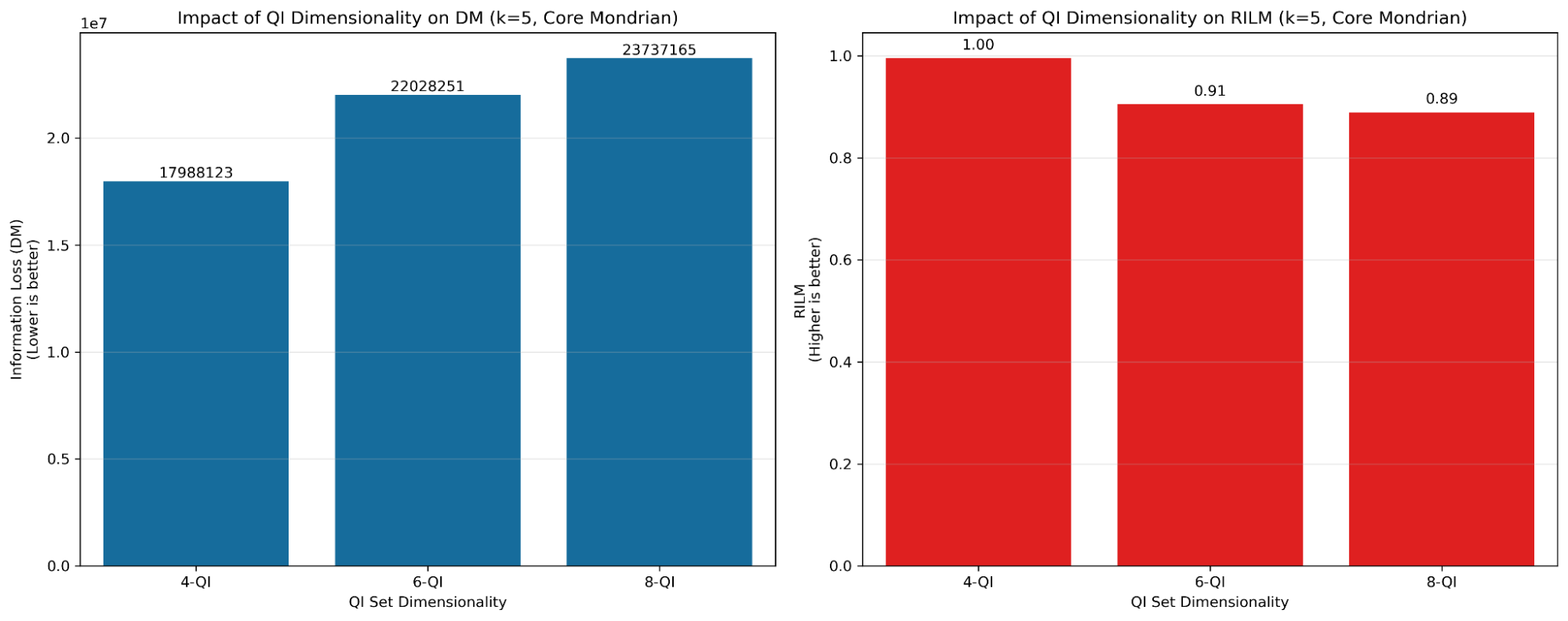}
\caption{Impact of QID dimensionality on information loss and utility}
\label{fig:qid-dimensionality-impact}
\end{figure}

{Figure~\ref{fig:qid-dimensionality-impact} shows that as the number of QIDs increases, information loss
grows substantially. The DM score increases from approximately 18m for 4
QIDs to approximately 24m for 8 QIDs. Correspondingly, the RILM utility
metric drops from nearly perfect 1.00 to 0.94. This highlights that
adding more QIDs makes it significantly harder to form anonymous groups
without extensive generalization.}

{}

\subsubsection{Effectiveness of Dynamic Suppression Management}

{Core Mondrian\textquotesingle s hybrid execution model enables dynamic
suppression budget management, particularly within its recursive
processing mode. Using the 6-QID set, we tested the impact of the
\texttt{recursive\_partition\_size\_cutoff}~parameter, which controls the
threshold for switching between recursive and queue-based processing.}

{}

\begin{figure}[h]
\centering
\includegraphics[width=0.8\textwidth,keepaspectratio]{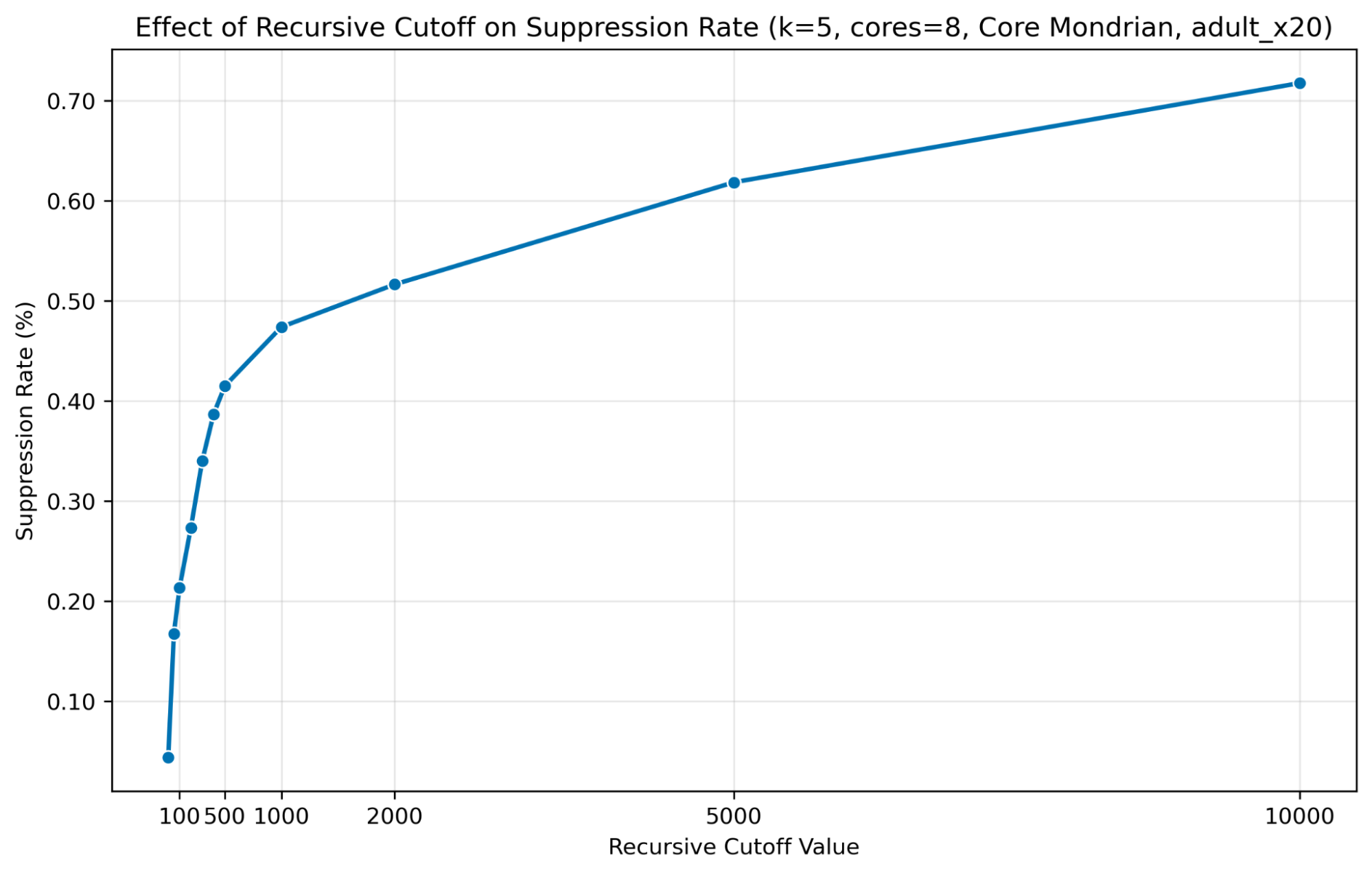}
\caption{Suppression rate decreases as recursive cutoff increases}
\label{fig:suppression-vs-cutoff}
\end{figure}

{Figure~\ref{fig:suppression-vs-cutoff} demonstrates that as the recursive cutoff increases, the
suppression rate drops sharply from over 0.12\% to near zero. A higher
cutoff means more partitions are handled by the recursive engine, which
can backtrack to find cuts that avoid suppression. In contrast, the
queue-based system for large partitions uses a fixed budget allocation.
This shows the recursive component\textquotesingle s capability to
preserve records that might otherwise require suppression. The plot
shows an optimal cutoff (elbow) value around \textbf{500-1000}. We use a
recursive cutoff of \textbf{1000}~in production.}

\subsection{\texorpdfstring{{Runtime Performance}}{Runtime Performance}}\label{h.v3lg5fkmyh5v}

{In production environments like Project Lighthouse, runtime performance
is as critical as data utility.}

{}

\subsubsection{Execution Time Comparison}

{We compared the runtime of Core Mondrian (on 1 and 8 cores) against the
single-threaded Original Mondrian across different }{k}{~values for a
4-QID set, with Dynamic Breakout disabled.}

{}

\begin{figure}[h]
\centering
\includegraphics[width=0.8\textwidth,keepaspectratio]{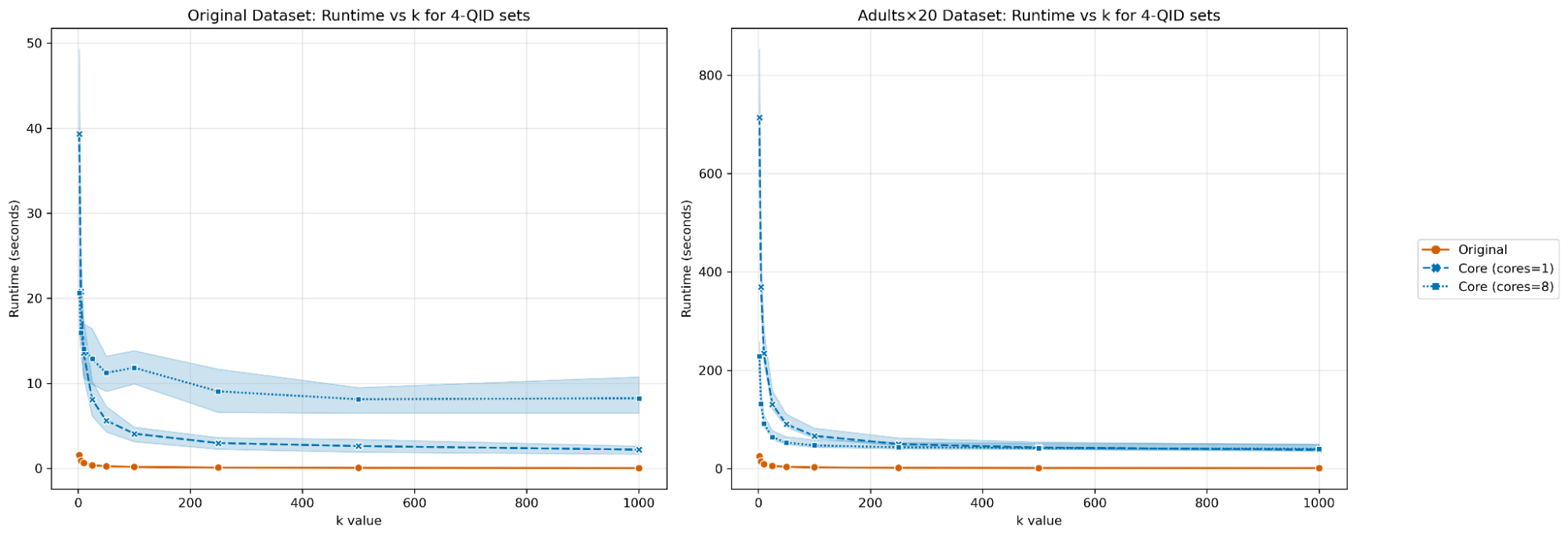}
\caption{Runtime comparison showing Core Mondrian's quality-performance tradeoff}
\label{fig:runtime-comparison}
\end{figure}

{As shown in Figure~\ref{fig:runtime-comparison}, Core Mondrian requires more computational time
than Original Mondrian, reflecting the additional work performed to
achieve higher quality anonymization. While Original Mondrian is faster,
this comes at a significant cost to output quality---as we demonstrated
earlier in Figure~\ref{fig:information-loss-vs-k}, Original Mondrian produces substantially higher
information loss (DM) and lower utility (RILM) for 4-QID sets across all
k values. Interestingly, the parallelization effects vary with dataset
size and k value: for smaller datasets (left plot), single-core Core
Mondrian (Core 1) is superior to eight-core (Core 8) for most k values,
likely due to parallelization overhead. However, for the larger dataset
and k values below 500, eight-core processing provides clear runtime
advantages. This pattern shows different performance characteristics
based on dataset size and k-values.}

{}

\subsubsection{Parallel Processing Scalability}

{To assess the effectiveness of our parallel framework, we measured
runtime and speedup on a large dataset (ADULT x20) using a comprehensive
6-QID set (}{age}{, }{marital-status}{, }{occupation}{,
}{native-country}{, }{sex}{, }{workclass}{) while varying the number of
processor cores, with Dynamic Breakout disabled.}

{}

\begin{figure}[h]
\centering
\includegraphics[width=0.8\textwidth,keepaspectratio]{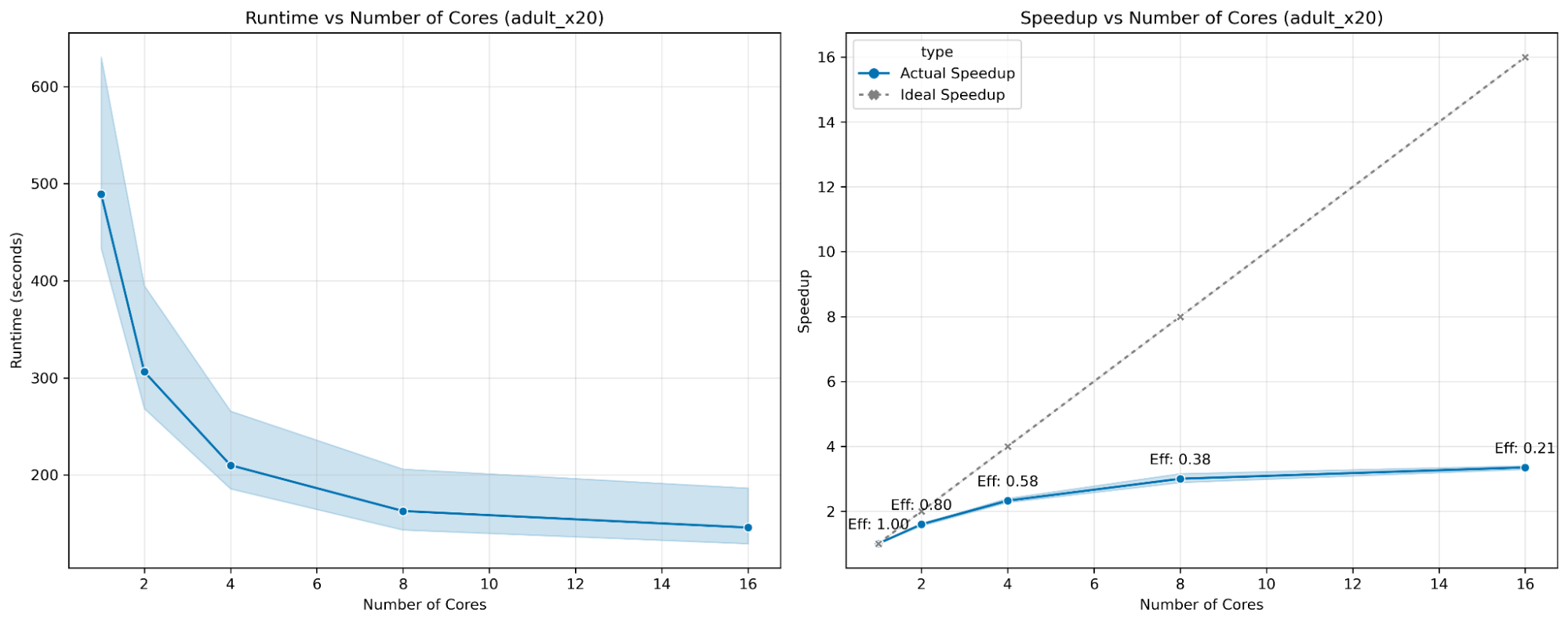}
\caption{Parallel processing scalability with up to 4× speedup}
\label{fig:parallel-scalability}
\end{figure}

{Figure~\ref{fig:parallel-scalability} shows that runtime decreases consistently as the number of
cores increases. The right panel plots the corresponding speedup
relative to a single-core run. While the actual speedup is sub-linear
and diverges from the ideal, this is expected due to parallelization
overhead and the inherently sequential portions of the algorithm.
Nonetheless, the results confirm that Core Mondrian scales effectively,
achieving a speedup of over 4x with 8 cores and demonstrating high
parallel efficiency. This scaling efficiency with a 6-QID set is
particularly important for practical applications where multiple
attributes must be protected simultaneously.}

{}

\subsubsection{Impact of Hybrid Execution Model on Runtime}

{The }{\texttt{recursive\_partition\_size\_cutoff}}{~parameter of the hybrid
execution model not only affects suppression (Figure~\ref{fig:suppression-vs-cutoff}) but also has a
significant impact on runtime. This analysis also used the 6-QID set to
provide direct comparison with the suppression results (Figure~\ref{fig:suppression-vs-cutoff}).}

{}

\begin{figure}[h]
\centering
\includegraphics[width=0.8\textwidth,keepaspectratio]{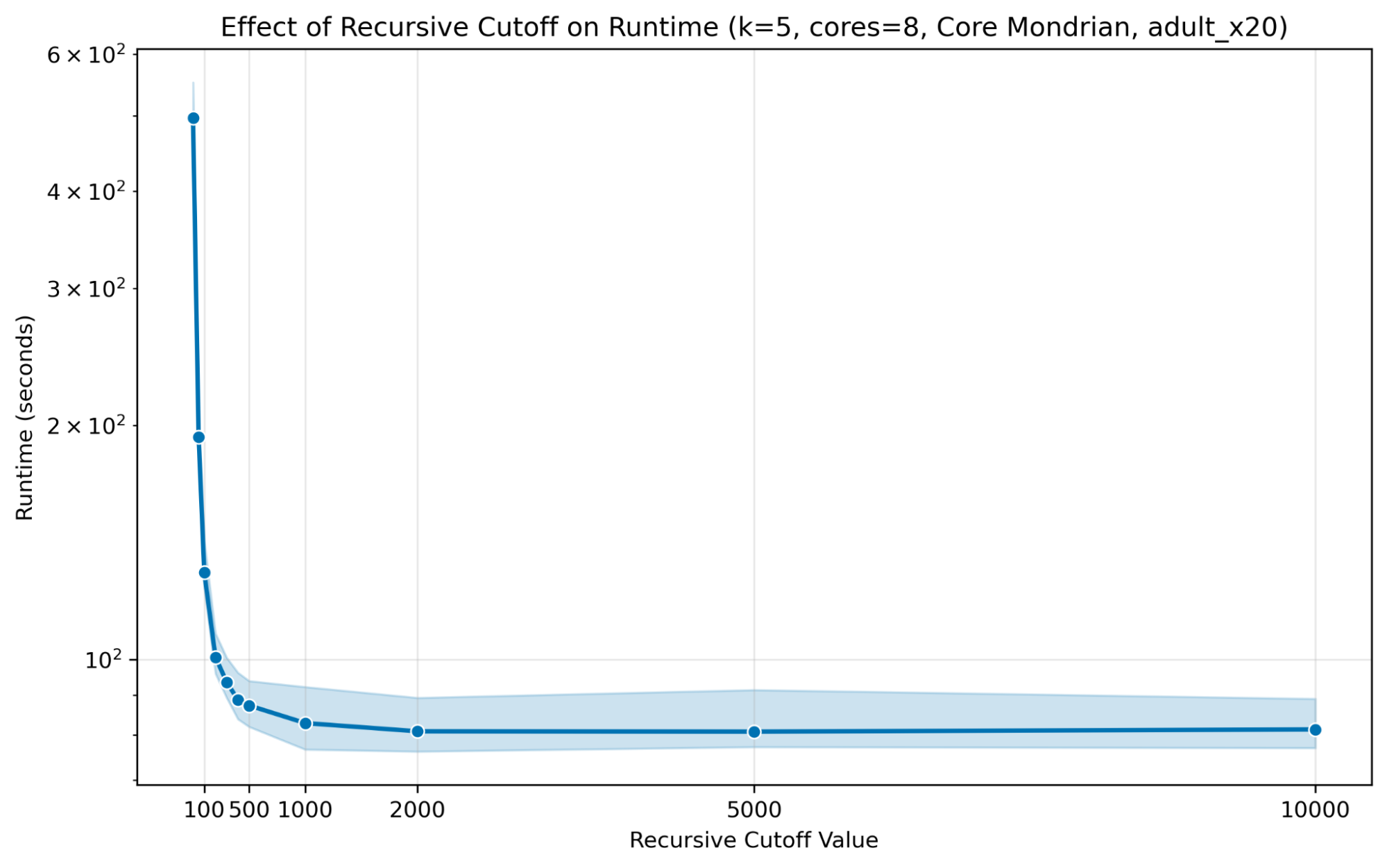}
\caption{Runtime vs recursive cutoff showing optimal threshold}
\label{fig:runtime-vs-cutoff}
\end{figure}

{Figure~\ref{fig:runtime-vs-cutoff} reveals a U-shaped relationship between the recursive cutoff
and runtime. At very low cutoffs, performance suffers due to the high
overhead of scheduling and managing a large number of very small tasks
in the parallel queue. The plot shows an optimal cutoff (elbow) value
around \textbf{500-1000}. We use a recursive cutoff of \textbf{1000}~in
production.}

\subsection{\texorpdfstring{{Cut Funnel Example}}{Cut Funnel Example}}\label{h.qsik29f2azmy}

\begin{figure}[h]
\centering
\includegraphics[width=0.8\textwidth,keepaspectratio]{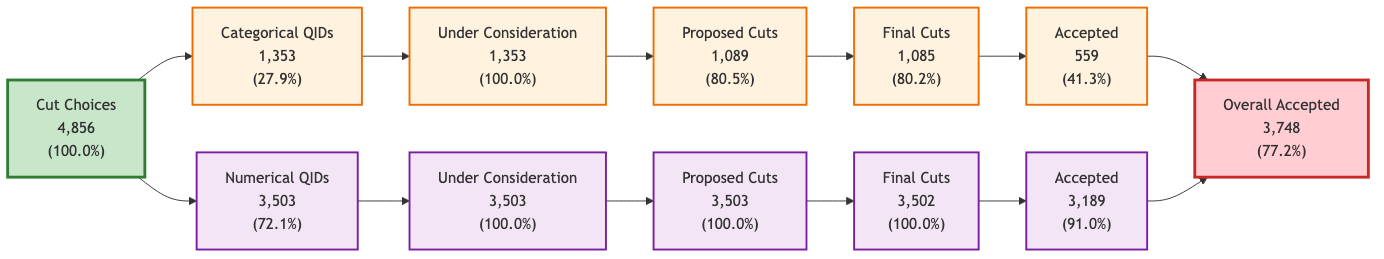}
\caption{Example cut funnel statistics for categorical (orange, top row) and numerical (purple, bottom row)}
\label{fig:cut-funnel-example}
\end{figure}

{To illustrate the cut funnel behavior (described at a high-level in
Section 3.3), we present an example of stagewise reduction of cut
candidates on the ADULT 6-QID dataset. Figure~\ref{fig:cut-funnel-example} shows the number of
candidate cuts present at each funnel stage, separately for categorical
and numerical QIDs (corresponding colors as indicated in the figure
legend).}

{}

{Key observations from this example:}

{}

\begin{itemize}
\tightlist
\item
  {There are many more \textbf{numerical than categorical cut choices}; one
  significant reason for this is that we are using "flat" generalization
  hierarchies whereby the root is "*" and the second layer are the leaf
  nodes (each unique categorical value).}
\item
  {The \textbf{acceptance rate is much lower for categorical than numerical
  cuts}---due to discarding categorical cuts that would violate the
  local suppression budget.}
\item
  {This example also demonstrates how \textbf{funnel statistics can enable
  empirical diagnosis}~of algorithmic bottlenecks and optimization
  opportunities.}
\end{itemize}

\subsection{\texorpdfstring{{Scalability Analysis}}{Scalability Analysis}}\label{h.8zoemrj1lz5i}

{Finally, we evaluate Core Mondrian\textquotesingle s ability to handle
large-scale data. We measured runtime and information loss as the dataset size was scaled
from its original size up to a \textbf{20x}~replication (\textasciitilde1M
records), using the 6-QID set with 8 cores and k=5.}

{}

\begin{figure}[h]
\centering
\includegraphics[width=0.8\textwidth,keepaspectratio]{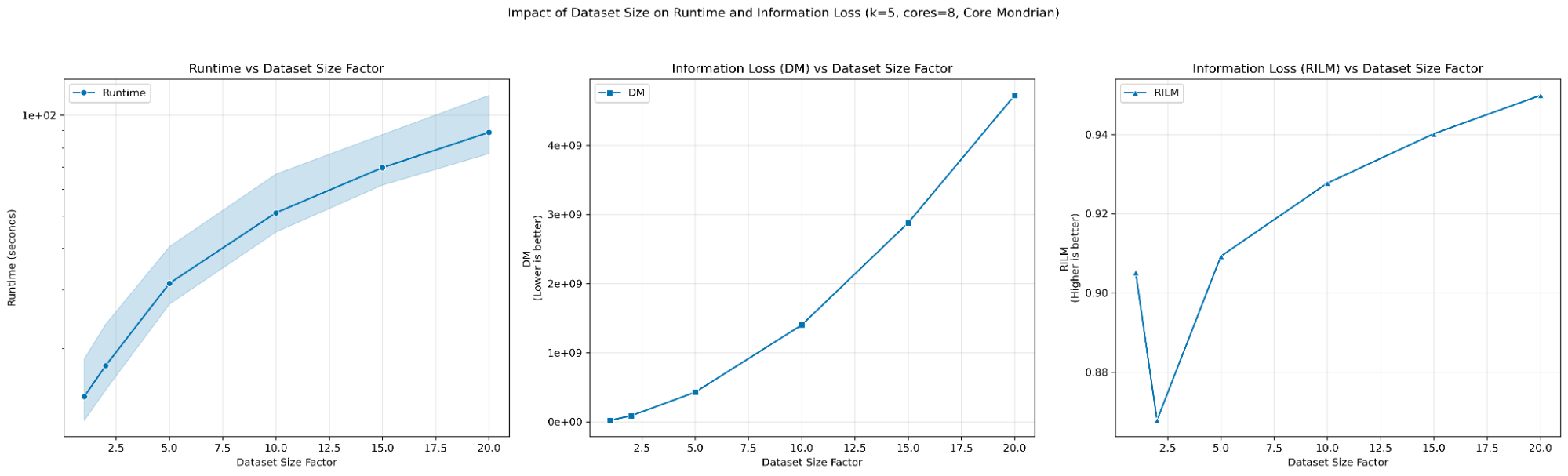}
\caption{Scalability analysis showing super-linear runtime growth}
\label{fig:scalability-analysis}
\end{figure}

{Figure~\ref{fig:scalability-analysis} demonstrates that runtime scales super-linearly with the
number of records, which is characteristic of partition-based
anonymization algorithms. Information loss, as measured by DM, also
increases steadily with dataset size, while utility (RILM) shows a
slight decline at first but then steadily increases with dataset size.}

{}

{These experimental results demonstrate Core Mondrian\textquotesingle s
effectiveness across multiple dimensions of performance, utility, and
scalability, confirming the value of our architectural and algorithmic
contributions.}

\section{\texorpdfstring{{Conclusion and Future Work}}{Conclusion and Future Work}}\label{h.surgmi29j99j}

{Core Mondrian extends the Mondrian k-anonymity algorithm with four key
contributions: extensible strategy-based architecture, hybrid parallel
execution, and utility-preserving enhancements (NaN-pattern
pre-partitioning and dynamic suppression budget management). Experiments
using UCI ADULT datasets demonstrate that Core Mondrian achieves lower
Discernibility Metric scores and higher RILM utility scores than
Original Mondrian for the tested numeric quasi-identifier
configurations. While Core Mondrian requires more computational time,
this runtime cost enables significantly higher quality anonymization.
Parallel processing delivers up to 4× speedup for Core Mondrian itself
compared to single-threaded execution. Dynamic suppression budget
management demonstrates clear tradeoffs between suppression and runtime
performance through backtracking mechanisms. We show that the system
scales to large datasets while maintaining deterministic, auditable
outputs essential for production analytics.}

{}

{Future work includes extending k-anonymity to scenarios where
individual users contribute \textbf{multiple records to the dataset}. The
modular Strategy Pattern architecture enables such extensions to
group-based privacy constraints. Additionally, \textbf{GPU acceleration}~of
parallel operations using cuDF for statistical computations while
maintaining CPU-based partitioning coordination could extend
applicability to larger datasets.}

\bibliography{references}

\end{document}